\documentclass[fleqn,twoside]{article}
\begin{document}
\begin{center}
\begin{large}
SPIN FRUSTRATION IN AN EXACTLY SOLVABLE ISING-HEISENBERG
DIAMOND CHAIN \\
\end{large}
\begin{large}
Michal Ja\v{s}\v{c}ur and Jozef Stre\v{c}ka \\
\end{large}
Department of Theoretical Physics and Astrophysics, Faculty of Science, \\
P.J. \v{S}af\'arik University, Moyzesova 16, 041 54 Ko\v{s}ice, Slovak Republic \\
E-mail: jascur@kosice.upjs.sk, jozkos@pobox.sk
\end{center}

\begin{abstract}
Exact results of a mixed spin-1/2 and spin-1  Ising-Heisenberg diamond chain
are obtained applying a generalized decoration-iteration transformation.
Depending on the values of external field, exchange parameters and anisotropy
there exist six different phases in  the ground state of the system. The
magnetic order of these phases is briefly described in this work.
\end{abstract}

During last two decades low-dimensinal magnetic materials have become important
subject in the research of magnetism.  Among systems exhibiting interesting
quantum magnetic phenomena one should mention those having a diamond-chain
structure that have been recently intensively studied both theoretically and
experimentally  (see \cite{Okamoto} and references therein). In order to
describe these materials, the quantum anisotropic Heisenberg model has been
frequently adopted. Though this model proves to be very useful, it brings
mathematical complexities preventing one to obtain exact results even for 1D
systems. For this reason, the main purpose of this work is to introduce a
simplified version of the Heisenberg model which is exactly solvable for the
diamond chain.  We will investigate here the diamond chain consisting of Ising- (gray circles) and Heisenberg-type atoms (black circles), as it is depicted in Fig. 1. In order to obtain exact expression for the partition function of the model, we  express the total Hamiltonian $\hat {\cal H}$ as a sum of the bond Hamiltonians, i.e.
$\hat {\cal H} = \sum_{k=1}^{N} \hat {\cal H}_k$, where
$N$ denotes the total number of the Ising-type atoms  and the bond Hamiltonian $\hat {\cal H}_k$ is given by (see Fig. 1.)
\begin{eqnarray}
\nonumber
\hat {\cal H}_k &=&
            J_1 (\hat S_{k1}^z + \hat S_{k2}^z)
               (\hat \mu_{k1}^z + \hat \mu_{k2}^z) \nonumber \\
&+& J_2 \bigl [\Delta (\hat S_{k1}^x \hat S_{k2}^x + \hat S_{k1}^y \hat S_{k2}^y
)
       + \hat S_{k1}^z \hat S_{k2}^z \bigr ]  \nonumber  \\
            &-& H_A (\hat \mu_{k1}^z + \hat \mu_{k2}^z)/2
            - H_B (\hat S_{k1}^z + \hat S_{k2}^z)
\label{eq1}
\end{eqnarray}
In this equation,   $\hat S_{k\alpha}^{\gamma}$  and $\hat \mu_{k\alpha}^z$
($\gamma = x, y,z; \; \alpha = 1, 2$) represent relevant components of the
standard spin-1 and spin-1/2 operators, respectively,  the exchange interaction
$J_1$ couples  the nearest-neighbouring  Ising and Heisenberg atoms,  $J_2$
couples nearest-neighbouring Heisenberg pairs and $\Delta$ represents the XXZ
anisotropy parameter. Finally, the terms including $H_A$ nad $H_B$ describe
interactions of relevant atoms with an external magnetic field. Next, introducing a generalized decoration-iteration transformation [2, 3] and utilizing the commutability of the bond Hamiltonians,  we easily find the following relation
\begin{equation}
{\cal Z}(\beta, J_1, J_2, \Delta, H_A, H_B)  = A^{N}{\cal Z}_0(\beta, R, H_0),
\label{eq2}
\end{equation}
where ${\cal Z} $ represents the partition function of our model and
${\cal Z}_0$ is the well-known partition function of a simple spin-1/2 Ising
linear chain and $\beta = 1/(k_B T)$.
For unknown transformation parameters $A$, $R$ and $H_0$, we obtain relatively
simple expressions [2, 3]:
\begin{eqnarray}
   A^{4} \!\!\!\!&=&\!\!\!\! (W_0 + W_1)(W_0 + W_2 )(W_0 + W_3)^{2}                             \\
 \beta R \!\!\!\!&=& \!\!\!\! \ln{\frac{(W_0 + W_1)(W_0 + W_2)}{(W_0 + W_3)^{2}}}       \\
 \beta H_0 \!\!\!\!&=&\!\!\!\! \beta H_A - \ln{(W_0 + W_1)} + \ln{(W_0 + W_2)}          \\
                W_0 \!\!\!\!&=&\!\!\!\! \exp(t_2) + f(t_2/2, a, 0, 0) - 4                        \\
                W_1 \!\!\!\!&=&\!\!\!\! f(-t_2, 2t_1 + 2 h_B, t_1 + h_B, t_2 \Delta)           \\
                W_2 \!\!\!\!&=&\!\!\!\! f(-t_2, 2t_1 - 2 h_B, t_1 - h_B, t_2 \Delta)           \\
                W_3 \!\!\!\!&=&\!\!\!\! f(-t_2, 2 h_B,  h_B, t_2 \Delta)
\label{eq3}
\end{eqnarray}
with $a = \sqrt{1 + 8 \Delta^2}$, $t_1 = \beta J_1,\; t_2 = \beta J_2,\; h_B = \beta H_B $, and
$f(x_1, x_2, x_3, x_4) = 2e^{x_1}\cosh x_2 + 4\cosh x_3\cosh x_4.$
Eqs. (2)-(9) enable one to calculate all physical quantities of interest.

In order to illustrate some interesting properties of the system, we have
depicted the ground-state phase  diagram in the $J_2-H$ space for the case of
the antiferromagnetic isotropic model i.e., $J_1>0$, $J_2>0$, $H_A = H_B = H$
and $\Delta = 1$. To describe different phases, we have calculated and analysed
the reduced magnetization of the Ising ($m_i^z$) and Heisenberg sublattice (
$m_h^z$) and various nearest-neighbour correlation functions ($C_{hh}^{xx}$,
$C_{hh}^{zz}$, $C_{ih}^{zz}$ and $C_{ii}^{zz}$). All these quantities are defined in the standard way, for instance,
$
m_i^z  \equiv \langle \hat \mu_{k\alpha}^z \rangle, \;
m_h^z  \equiv \langle \hat S_{k\alpha}^z  \rangle,\;
C_{hh}^{xx} \equiv \langle \hat S_{k1}^x \hat S_{k2}^x \rangle,
$
where $\langle ... \rangle$ means standard canonical average,
subscripts $i (h)$ reffer to the Ising (Heisenberg) atoms
and superscripts represent spatial components of the spin.
As one can see from Fig.2, there exist six different phases separated
by first-order phase transition boundaries. With the help  of  above introduced
quantities we can characterize these phases as follows:\\
Ordered phase 1 (OP$_1$):\\
$m_i^z=-1/2$, $m_h^z=1$, $C_{hh}^{zz} = 1$, $C_{hh}^{xx} = 0$,
$C_{ih}^{zz} = -1/2$, $C_{ii}^{zz} = 1/4$;\\
Ordered phase 2 (OP$_2$):\\
$m_i^z=-1/2$, $m_h^z=1/2$, $C_{hh}^{zz} = 0$, $C_{hh}^{xx} = -1/2$,
$C_{ih}^{zz} = -1/4$, $C_{ii}^{zz} = 1/4$;\\
Ordered phase 3 (OP$_3$):\\
$m_i^z=1/2$, $m_h^z=0$, $C_{hh}^{zz} = -2/3$, $C_{hh}^{xx} = -2/3$,
$C_{ih}^{zz} = 0$, $C_{ii}^{zz} = 1/4$;\\
Ordered phase 4 (OP$_4$):\\
$m_i^z=1/2$, $m_h^z=1/2$, $C_{hh}^{zz} = 0$, $C_{hh}^{xx} = -1/2$,
$C_{ih}^{zz} = 1/4$, $C_{ii}^{zz} = 1/4$;\\
Saturated paramagnetic phase (SP):\\
$m_i^z=1/2$, $m_h^z=1$, $C_{hh}^{zz} = 1$, $C_{hh}^{xx} = 0$,
$C_{ih}^{zz} = 1/4$, $C_{ii}^{zz} = 1/4$;\\
Frustrated phase (FP):\\
$m_i^z=0$, $m_h^z=0$, $C_{hh}^{zz} = -2/3$, $C_{hh}^{xx} =-2/3 $,
$C_{ih}^{zz} = 0$, $C_{ii}^{zz} = 0$.\\
As one can see the OP$_1$ and SP represent standard phases that are usually
observed also in the pure Ising models. On the other hand, the remaining  four phases exhibit apparently a nontrivial spin ordering related to strong quantum
fluctuations in the system. Due to this effect magnetic properties of the OP$_2$
and OP$_4$ can be explained within the valence-bond-solid picture as suggested in \cite{Strecka2}. Moreover, the stepwise magnetization curves exhibit plateaux at relevant ordered phases OP$_1$, OP$_2$, OP$_3$ or OP$_4$.
 In addition to the above mentioned phases, there appears for
$H/J = 0 $ and $J_2>J_1$ an interesting frustrated phase (FP) having
the nonzero entropy, namely, $S/N k_B = \ln 2$. Even without presenting
further details here, it is clear that the system under consideration is very
interesting and it will exhibit nontrivial  properties also at finite
temperatures.

{\it Acknowledgment}: This work was supported under grants
                      APVT 20-009902 and VEGA 1/9034/02
\newline
\newline
\newline
\newline
FIGURE CAPTIONS:
\newline
\newline
Fig.1: Part of the of the mixed-spin diamond chain.
Gray circles represent spin-1/2 Ising atoms and black ones
spin-1 Heisenberg atoms. The ellipse demarcates $k$th bond (see text)
\newline
\newline
Fig.2: Ground-state phase diagram of the Ising-Heisenberg diamond chain
in the $J_2-H$ space. OP$_1$, OP$_2$, OP$_3$, OP$_4$, SP and FP denote
different  phases described in the text


\begin{thebibliography}{4}
\bibitem{Okamoto} K. Okamoto and Y. Ichikawa,
                  J. Phys. Chem. Solids 63 (2002) 1575.
\bibitem{Fisher} M. E. Fisher, Phys. Rev. 113 (1959) 969.
\bibitem{Strecka1} J. Stre\v{c}ka and M. Ja\v{s}\v{c}ur, Phys. Rev. B 66 (2002)
174715.
\bibitem{Strecka2} J. Stre\v{c}ka and M. Ja\v{s}\v{c}ur, Phys. Stat. Sol. (b) 233 (2002) R12.
\end{thebibliography}
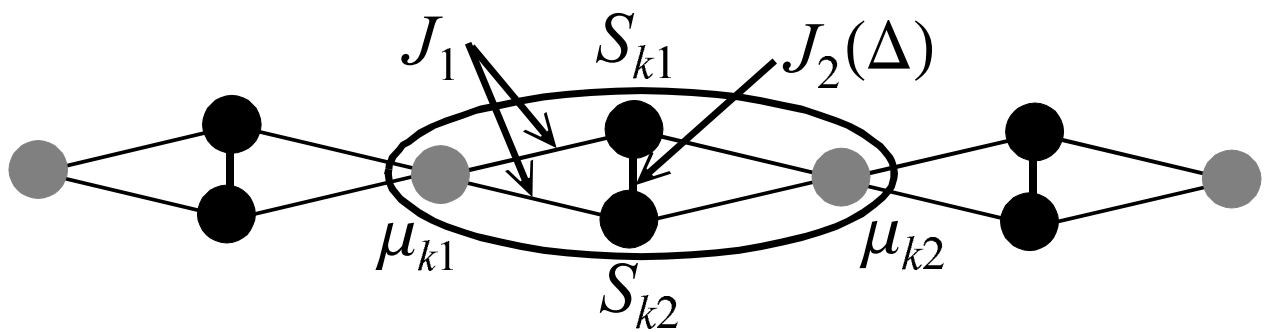
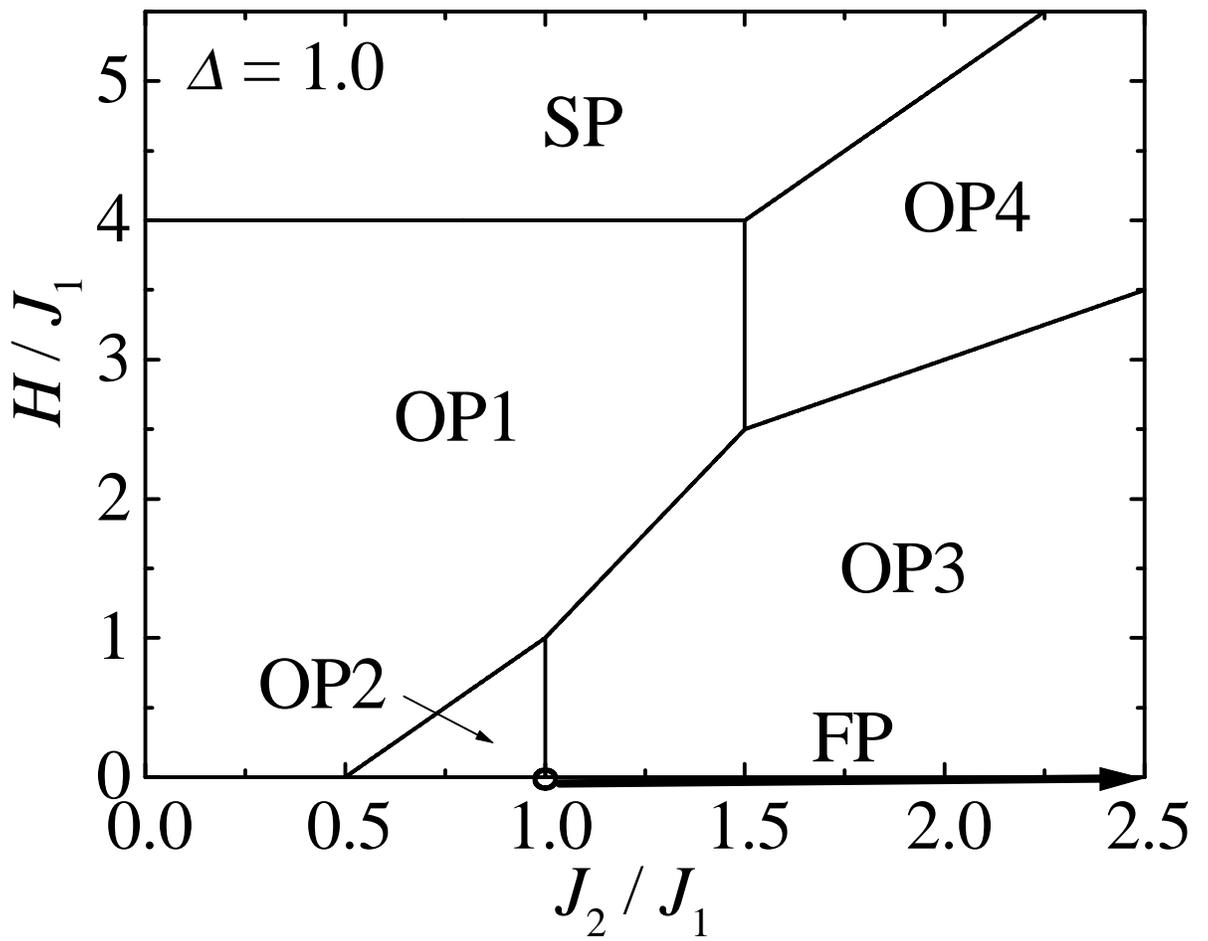
\end{document}